\documentclass[aps,pra,reprint,superscriptaddress,floatfix]{revtex4-2}

\usepackage{graphicx, textcomp, graphics, epsfig}
\graphicspath{ {./figures/} }
\usepackage{amsmath,amssymb,amsthm,wasysym, latexsym, braket}

\usepackage{xcolor}
\usepackage{color}
\definecolor{LinkColor}{rgb}{0.256,0.439,0.588}
\usepackage{bm}      	
\usepackage{longtable}
\usepackage{siunitx}

\usepackage{hyperref}
\hypersetup{
    colorlinks=true,
    citecolor=LinkColor,
    linkcolor=LinkColor,
    urlcolor=LinkColor
}
\hypersetup{linktocpage}
\usepackage[nameinlink,capitalise]{cleveref}

\newcommand{\blank}{\,&}

\newcommand{\fig}[1]
{Fig.~\ref{fig:#1}}
\newcommand{\eq}[1]
{Eq.~\eqref{eq:#1}}
\newcommand{\sect}[1]
{Sec.~\ref{sec:#1}}

\def\bmr{\bm{r}}

\def\wf{W_{i\mu}}
\def\wfr{\wf(\bmr)}

\def\V#1{V_{#1}}
\def\Vint{\V{\mathrm{int}}(\bmr-\bm{r'})}
\def\Vtrap{\V{\mathrm{trap}}}
\def\Vtrapr{\Vtrap(\bmr)}
\def\Vtot{\V{\mathrm{total}}(\bmr)}
\def\VrR#1#2{\V{#1,#2}(\bmr-\bm{R}_#2)}

\begin{document}


\title{Hubbard parameters for programmable tweezer arrays}

\author{Hao-Tian Wei}
\email[]{htwei@rice.edu}
\affiliation{Department of Physics and Astronomy, Rice University, Houston, Texas 77005-1892, USA}
\affiliation{Rice Center for Quantum Materials, Rice University, Houston, Texas 77005-1892, USA}
\author{Eduardo Ibarra-Garc\'ia-Padilla}
\affiliation{Department of Physics and Astronomy, Rice University, Houston, Texas 77005-1892, USA}
\affiliation{Rice Center for Quantum Materials, Rice University, Houston, Texas 77005-1892, USA}
\affiliation{Department of Physics, University of California, Davis, California 95616, USA}
\affiliation{Department of Physics and Astronomy, San Jos\'e State University, San Jos\'e, California 95192, USA}
\author{Michael L. Wall}
\affiliation{The Johns Hopkins University Applied Physics Laboratory, Laurel, Maryland 20723, USA}
\author{Kaden R. A. Hazzard}
\affiliation{Department of Physics and Astronomy, Rice University, Houston, Texas 77005-1892, USA}
\affiliation{Rice Center for Quantum Materials, Rice University, Houston, Texas 77005-1892, USA}
\affiliation{Department of Physics, University of California, Davis, California 95616, USA}

\date{\today}

\begin{abstract}
The experimental realization of Fermi-Hubbard tweezer arrays
opens a new stage for engineering fermionic matter, where programmable lattice geometries and Hubbard model parameters are combined with single-site imaging. In order to use these versatile experimental Fermi-Hubbard models as quantum simulators, it is crucial to know the Hubbard parameters describing them. Here we develop methods to calculate the Hubbard model parameters of arbitrary two-dimensional lattice geometries: the tunneling $t$, on-site potential $V$, and interaction $U$, for multiple bands and for both fermions and bosons. We show several examples. One notable finding is that a finite array of equally strong and separated individual tweezer potentials actually sums to give a non-periodic total potential and thus spatially non-uniform Hubbard parameters. We demonstrate procedures to find trap configurations that equalize these parameters. More generally, these procedures solve the inverse problem of calculating Hubbard parameters: given desired Hubbard parameters, find trap configurations to realize them. These methods will be critical tools for using tunnel-coupled tweezer arrays.
\end{abstract}

\maketitle

\section{Introduction}
\def\teal#1{\textcolor{teal}{#1}}

The Fermi-Hubbard model has been widely studied over a half century as it captures a key feature of strongly correlated matter, the competition between kinetic and interaction energy in a lattice, which is relevant to almost all quantum materials. Although it is the simplest model for studying interacting fermions on a lattice, it displays rich physics, such as a metal-insulator crossover, antiferromagnetic order, strange metallicity, and potential $d$-wave superconductivity~\cite{imada1998,montorsi1992,Tasaki1998,Arovas2022,Qin2022,Hubbard_model}. Due to its richness, the Fermi-Hubbard model is of fundamental interest, and is studied in numerous quantum simulation platforms.

Ultracold atoms in optical lattices have been paradigmatic quantum simulators of Hubbard models. Numerous long-studied Fermi-Hubbard phenomena have been observed and explored using optical lattice experiments~\cite{Esslingerreview,QuantSim,Tarruell_review,Bloch2012,Gross2017}. Recently, quantum gas microscopes capable of resolving single lattice sites have further extended the capability of quantum control and quantum simulation~\cite{GrossBakr_review,Bohrdt2021,Altman2021}.

Nevertheless, optical lattice Hubbard models have important limitations, most apparently that they are restricted to periodic potentials since they are formed by interfering lasers. Additionally, optical lattice experiments have yet to reach temperatures deep into the regime characteristic of 
antiferromagnetism or potential superconducting order \cite{Mazurenko2017,Taie2022}.  

Recently, atoms in tunnel-coupled optical tweezer arrays have provided a  platform for simulating Hubbard models with programmable one-dimensional (1D) \cite{BosonDoubleWell,FermionDoubleWell,Bergschneider2019,Becher2020,Spar2022,Florshaim2023} and two-dimensional (2D) \cite{Yan2022} geometries, and dramatically new capabilities for reaching lower temperatures and entropies. In particular, using near-unit filling and post-selection techniques to prepare low-entropy initial states for adiabatic preparation may allow experiments to access  previously inaccessible low-temperature phases~\cite{Yan2022,Spar2022}. 

However, to utilize tunnel-coupled tweezers as quantum simulators, one needs to know the programmed Hubbard model parameters: on-site potentials, tunneling rates, and interactions. Theory is necessary, as measuring or calibrating all of these parameters experimentally is challenging -- cross-effects between traps make independently measuring parameters difficult.  

Although in principle these parameters can be determined from the single-particle eigenstates, the calculations are significantly more complicated than for optical lattices. In both cases, one first determines the single-particle eigenstates and then  uses these to calculate localized Wannier functions from which Hubbard parameters are obtained. For tweezers, however, there are obstacles in both steps. This has restricted calculations to simple analytic treatments~\cite{HagueAnalyticParam2021};  although useful, this gives only rough order-of-magnitudes estimates, and it also  misses qualitative features, such as the correct scaling of the tunneling with respect to trap depth. 

The first obstacle is the challenge of finding the eigenstates. Tweezer arrays are non-separable, in contrast to many optical lattices, which  reduce to one-dimensional problems. Additionally, tweezers have no spatial periodicity, increasing the size of systems that must be considered. Despite these challenges, discrete variable representation (DVR) methods \cite{DVR,Nygaard2003,Nygaard2004,LittlejohnDVR2002,Fornberg1996,Colbert1992} that were first applied to two tunnel-coupled tweezers in Ref.~\cite{Wall2015} are able to efficiently find the eigenstates. DVR methods combine the exponential convergence of spectral methods with the sparsity of real-space finite-difference methods.

The second obstacle is that determining appropriately localized Wannier functions is more involved than for lattices, even after the single particle eigenbasis is found. For optical lattices, the Wannier functions are simply Fourier transforms of the eigenstates because of the periodicity of the lattice. While for only two identical tweezers, one can determine the Wannier functions from reflection symmetry~\cite{Wall2015}, for general tweezer arrays finding the unitary basis transform that gives localized Wannier functions are more involved. 

In this paper, we develop a method to compute the Hubbard parameters for tunnel-coupled tweezer arrays given the trap parameters. We illustrate this method on several 1D and 2D geometries and to multi-band models. The method works by combining the DVR method for single-particle eigenstates with Riemannian manifold optimization techniques to compute maximally localized Wannier functions that have been developed in condensed matter and quantum chemistry \cite{MLWF}. In the context of ultracold matter, similar methods were used to construct Wannier functions for 1D double-well and 2D honeycomb lattices \cite{Modugno1D2012,MLWF2D2013}, but for these periodic lattices Bloch eigenstates can be obtained by diagonalizing a single unit cell so DVR techniques are less necessary, and Wannier functions are translations of each other. There exists research on non-periodic quasicrystalline optical lattices as well \cite{GottlobQuasicrystal2023}. However, this work doesn't apply the more efficient DVR method, and the way it constructs Wannier functions may lose the maximal locality.

One important finding of our example calculations is that even when tweezers are nominally uniform -- the Gaussian beams have equal spacings, waists, and strengths -- the resulting Hubbard parameters vary spatially because the total trapping potential differs site-to-site due to the finite-size of the array.
This presents an obstacle to simulating translational invariant many-body systems, which have uniform Hubbard parameters.
We therefore propose and demonstrate techniques to equalize the Hubbard parameters.

The paper is organized as follows. \sect{HubbardParam} presents the method and demonstrates Hubbard parameter calculations in optical tweezer arrays. We then describe our methods and propose two experimental protocols to increase control of Hubbard parameters in \sect{equalizeHubbard}, and demonstrate these suffice to achieve spatially uniform Hubbard parameters. 
\sect{conclusion} concludes.
Our source code is publicly available on GitHub~\cite{HubbardTweezer}. 

\section{Hubbard parameter calculations}
\label{sec:HubbardParam}

This section shows the methods and results of our Hubbard parameter calculations. 
\sect{effModel} presents the theoretical description of tunnel-coupled tweezer arrays and outlines our approach to calculating Hubbard parameters. 
\sect{DVR} describes the DVR method to obtain single-particle eigenstates. \sect{MLWF} describes how to construct the maximally-localized Wannier functions (MLWFs) from which the Hubbard parameters are calculated.
\sect{HubbardParamResults} then calculates and discusses the Hubbard parameters in various 1D and 2D lattice geometries. This exhibits the power of the tweezer platform and the efficiency and flexibility of our algorithm.


\subsection{Tweezer arrays and outline of obtaining Hubbard parameters}
\label{sec:effModel}

\begin{figure}[t]
    \centering
    \includegraphics[width=0.5\textwidth]{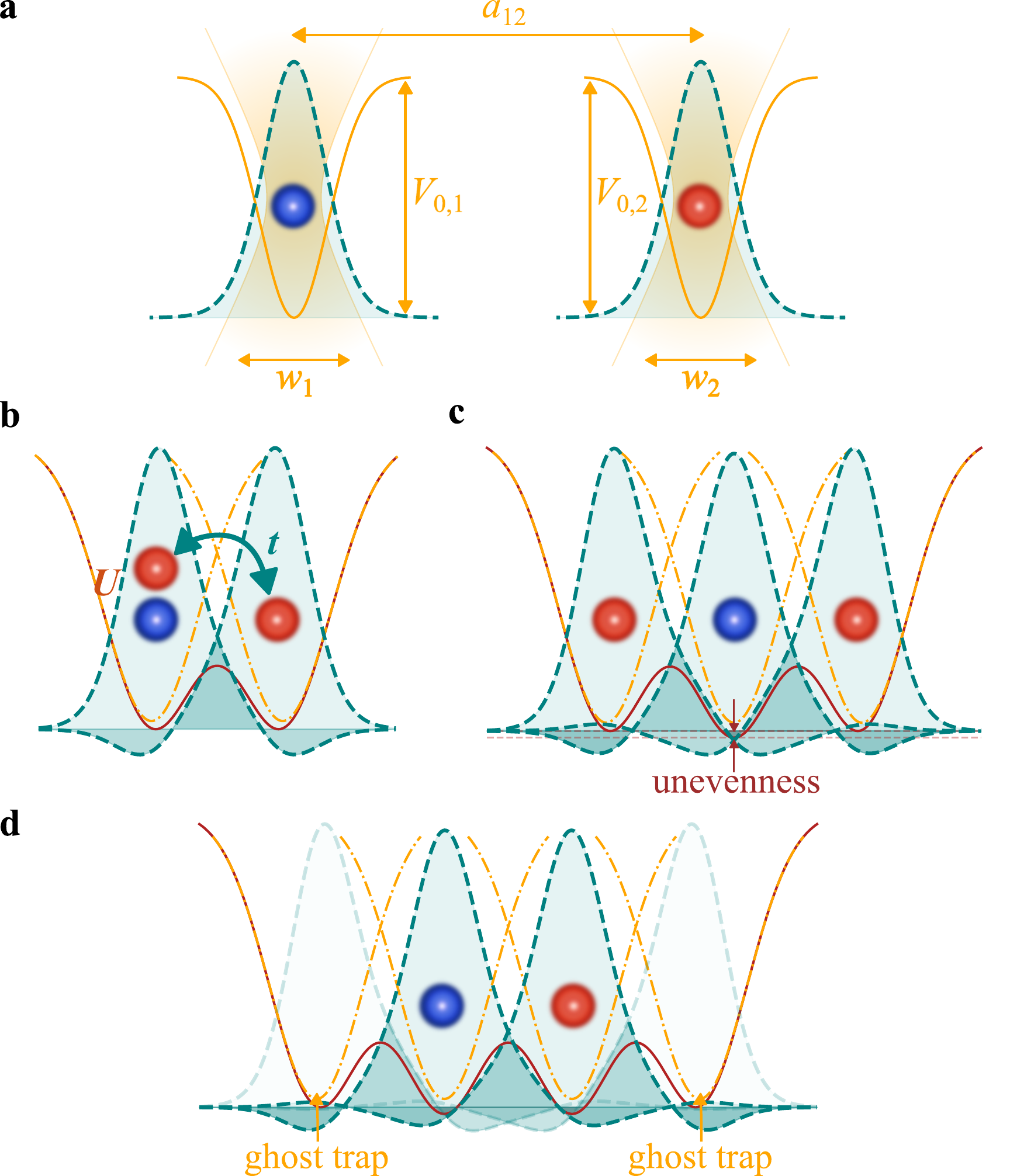}
    \caption{Hubbard model realized in an optical tweezer array. (a) Tweezers are brought together in an array, with tunable trap depths $V_{0,i}$, spacings $a_{ij}$, and waists $w_i$, where indices index tweezers. Throughout, we treat the waists $w_i$ as fixed unless otherwise noted, since they  are not tunable in current experiments \cite{Spar2022,Yan2022}. Only in \sect{waist} do we present a speculative proposal for trap engineering, where we allow  the waists to be tunable.  (b) Under suitable conditions, the array is described by a low-energy effective Hubbard model with tunnelings $t_{ij}$ and interactions $U_i$. (c) Due to cross-talk, setting distances between traps and trap depths equal gives potentials, interactions, and tunneling rates that are inhomogeneous: the existence of the right trap changes the energy barrier between the left and center traps, for example. (d) Two methods to equalize Hubbard parameters are to adjust trap waists (in addition to locations and depths), as is shown in (a), or to add ``ghost traps" as   in (d).}
    \label{fig:Tweezer}
\end{figure}

The optical tweezer array is made of tightly-focused Gaussian lasers, each of which induces an attractive potential  [\fig{Tweezer}(a)],
\begin{equation}
    \Vtrapr = \frac{-V_0}{1+\frac{z^2}{z^2_R}} \exp\left[\sum_{\xi=x,y}\frac{-2\xi^2}{\left(1+\frac{z^2}{z^2_{R,\xi}}\right)w^2_\xi}\right], 
\end{equation}
where $\bmr=(x,y,z)$ is the Cartesian coordinate vector, and $\xi$ goes over $x$ and $y$ transversal coordinates only, $V_0$ is the trap depth, $w_x$ ($w_y$) and $z_{R,x}$ ($z_{R,y}$) is the trap waist and Rayleigh range in $x$ ($y$) direction. $z_R$ is the ``effective'' Rayleigh range, defined by $\frac{1}{z^2_R} = \frac{1}{2}\left(\frac{1}{z^2_{R,x}} + \frac{1}{z^2_{R,y}}\right)$. The trapping laser beams' locations and depths are programmable via deflections from a driven acousto-optic modulator (AOM), or potentially phase masking by a spatial-light modulator (SLM).

Bringing traps close enough together results in coherent tunnel-coupling, and in the right regime (described in what follows) gives a system described by a Fermi-Hubbard model [\fig{Tweezer}(b)]. The continuum Hamiltonian of atoms in the tweezer array is 
\begin{align}\label{eq:contmHamiltonian}
    H=\blank H_0 + H_\mathrm{int} \nonumber\\
    =\blank \int \! d^d\bmr \, \psi^\dagger(\bmr)\left[-\frac{\hbar^2}{2m}\nabla^2 + \Vtot \right]\psi(\bmr) \nonumber\\
    \blank+ \frac{1}{2} \int \! d^d\bmr \,\! d^d\bm{r'} \, \psi^\dagger(\bmr)\psi^\dagger(\bm{r'})
    \Vint \psi(\bm{r'})\psi(\bmr),
\end{align}
with total trapping potential
\begin{equation}
    \Vtot = \sum_i\VrR{\mathrm{trap}}{i}
\end{equation}
where $\bm{R}_i$ are the beam centers, $\hbar$ is the reduced Planck constant, $m$ is the atom mass, $\psi(\bm{r})$ [$\psi^\dagger(\bm{r})$] is the annihilation [creation]  operator, and $\Vint$ is the interaction, which in atomic systems, where the typical particle separation is large compared to the interaction range, is
\begin{equation}
    \Vint = \frac{4\pi\hbar^2a_s}{m}\delta(\bmr-\bm{r'})
\end{equation}
where $a_s$ is the $s$-wave scattering length. In principle, this interaction needs to be regularized, but for what we will do this unregularized form suffices since we will only use matrix elements in a finite basis of analytic functions.
Longer-ranged interactions, as occur in ultracold molecules \cite{Gorshkov2011,Yan2013,dipolarFermiMolecule2015} and Rydberg-dressed atoms \cite{DefenuLRI2021,JohnsonRydbergDress2010,RydbergHubbard2015}, can be incorporated into our theory straightforwardly.

Under suitable conditions, the continuum Hamiltonian, Eq.~\eqref{eq:contmHamiltonian} is equivalent to a Hubbard Hamiltonian. Specifically, when the interaction strength and temperature are weak compared to energy gaps to excited states out of the manifold that we will keep to describe the Hubbard limit (analogous to ``band gaps" in infinite periodic systems), and the number of particles is sufficiently low so that single-particle states above this gap are not occupied, one can project Eq.~\eqref{eq:contmHamiltonian} to the states necessary for an accurate description of the physics, which is often only a few bands or even a single band. Then Eq.~\eqref{eq:contmHamiltonian} is 
\begin{align}
    H_\mathrm{FH} =\blank
    - \sum_\mu\sum_{ij} t_{ij\mu} a^\dagger_{i\mu} a_{j\mu}
    + \sum_\mu\sum_{i} V_{i\mu} n_{i\mu} \nonumber\\
    \blank+ \sum_{\mu\nu\delta\sigma}\sum_{ijkl} U_{ijkl;\mu\nu\delta\sigma} a^\dagger_{i\mu}a^\dagger_{j\nu} a_{k\delta}a_{l\sigma}, \label{eq:FH}
\end{align}
where $\mu$, $\nu$, $\delta$ and $\sigma$ index the ``bands", $i$, $j$, $k$ and $l$ index lattice sites, and $a_{i\mu}$ ($a^\dagger_{i\mu}$) is the operator that annihilates (creates) an atom on an orbital $\wfr$, which is related to the field operators via the transformation
\begin{equation}
    \psi(\bmr)=\sum_{i\mu}\wfr a_{i\mu},
\end{equation}
and the Hubbard parameters are given by
\begin{align}
    & t_{ij\mu} = -\int \! d^d\bmr\, W^*_{i\mu}(\bmr)\left[-\frac{\hbar^2}{2m}\nabla^2 + \Vtot\right] W_{j\mu}(\bmr) \label{eq:t} \\
    & V_{i\mu} = \int \! d^d\bmr\, W^*_{i\mu}(\bmr) \left[-\frac{\hbar^2}{2m}\nabla^2 + \Vtot\right] W_{i\mu}(\bmr) \label{eq:V} \\
    & U_{ijkl;\mu\nu\delta\sigma} = \frac{4\pi\hbar^2a_s}{m} \int \! d^d\bmr\, 
    W^*_{i\mu}(\bmr) W^*_{j\nu}(\bmr) W_{k\delta}(\bmr) W_{l\sigma}(\bmr). \label{eq:U}
\end{align}

In general, the set of orbitals $\{W_{i\mu}(\bmr)\}$ can be an arbitrary single-particle basis spanning the truncated space. However, it is useful to choose $W_{i\mu}(\bmr)$ to be as localized as possible, (by some measure introduced later), referred to as maximally localized Wannier functions (MLWF), in which case the $t_{ij\mu}$ and $U_{ijkl;\mu\nu\delta\sigma}$ can be truncated by discarding matrix elements between sites far enough apart in real space. In sufficiently deep lattices, it is an excellent approximation to truncate $t_{ij\mu}$ to nearest neighbors and $U_{ijkl;\mu\nu\delta\sigma}$ to on-site interactions, in which case we will denote it $U_{i;\mu\nu\delta\sigma}$. We will often suppress the spatial indices if the parameters are uniform and suppress the band index in the single-band case.

\begin{figure}
    \centering
    \includegraphics[width=0.48\textwidth]{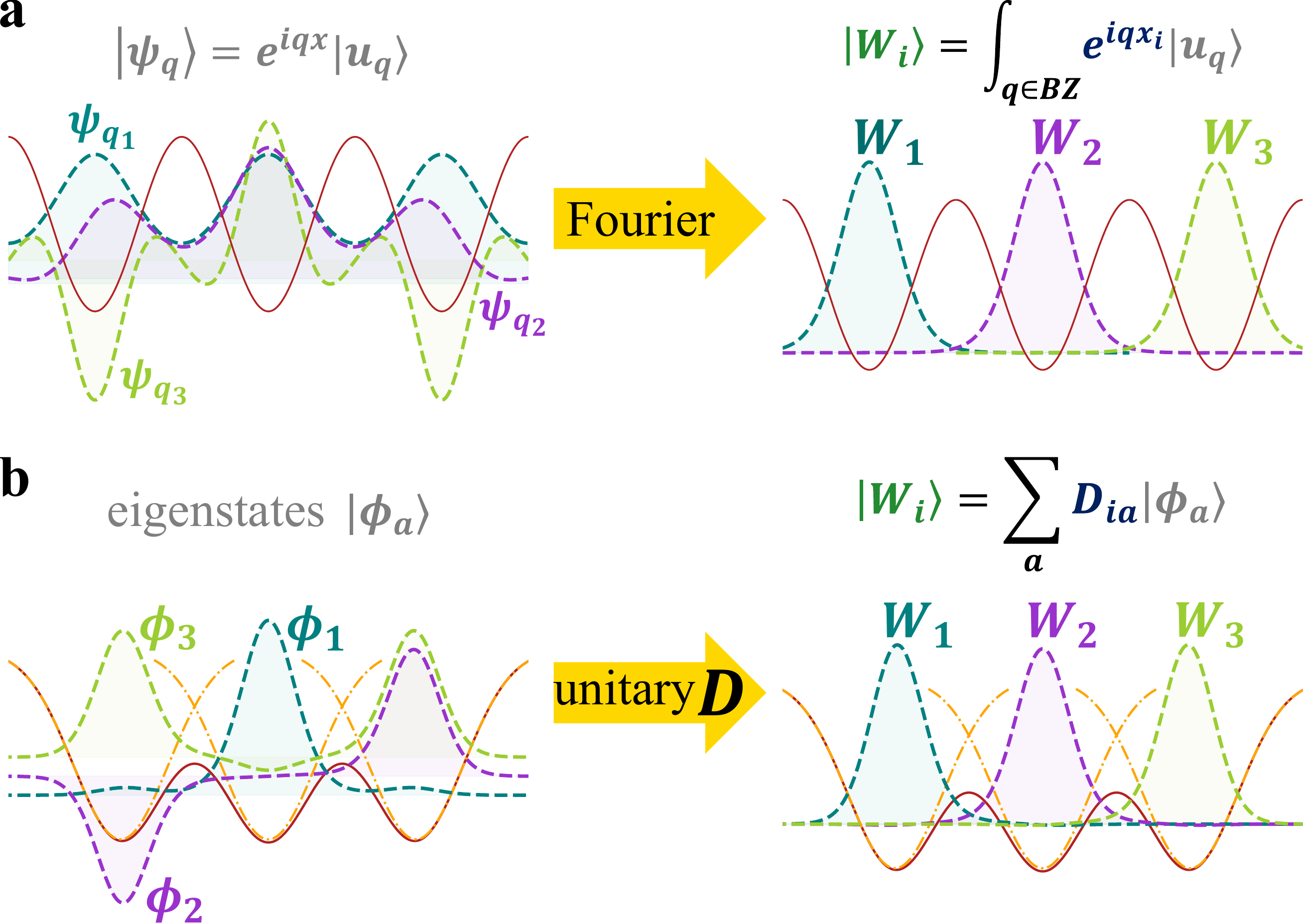}
    \caption{Maximally localized Wannier functions. (a) In a typical optical lattice the Wannier functions are Fourier transforms of the Bloch wavefunctions of the bands of interest (usually lowest energy), while (b) in tweezer arrays, the Wannier functions are unitary transforms $D$ of the relevant single-particle eigenstates (usually lowest energy), where $D$ is chosen to produce the maximally localized set of Wannier functions, as measured by the cost function Eq.~\eqref{eq:localization-cost}. Maximal localization ensures that the tunneling and interactions in the effective Hubbard model resulting from this basis can be truncated to short distances.}
    \label{fig:Wannier}
\end{figure}

In infinite periodic lattices, the MLWFs are simply Fourier transforms of the single-particle eigenstates (Bloch wavefunctions), as shown in \fig{Wannier}(a). The situation for optical tweezer arrays is more complicated, and more general unitary transformations from the single-particle eigenstates are required, illustrated in \fig{Wannier}(b)].  

Our method, detailed in the next two subsections, is, therefore, to (1) solve for the single-particle eigenstates, (2) find the MLWFs by finding the unitary transforms that give MLWFs in the eigenstate basis truncated to the desired low-energy bands, and (3) compute the Hubbard parameters with Eqs.~\eqref{eq:t},~\eqref{eq:V}, and~\eqref{eq:U}.


\subsection{Discrete variable representations}
\label{sec:DVR}

We employ the DVR method, following Ref.~\cite{Wall2015}, to calculate the single-particle eigenstates with the lowest energies.
The DVR method solves the Schr{\"o}dinger equation in a basis of position basis states at grid points $\bmr_n$ (inside some spatial region) projected into a low-momentum subspace. It maintains the exponential convergence of spectral methods and sparsity of methods that use finite-difference derivatives on real-space finite grids. All matrix elements are simple to obtain analytically. The sparse Hamiltonian can be diagonalized by the Lanczos algorithm to obtain the lowest energy eigenstates. Besides the benefits above, symmetries, such as spatial reflection or rotation, can be easily incorporated into the choice of DVR basis, to reduce the size of the matrix. The DVR method was shown to be an effective method for treating atoms in optical tweezers~\cite{Wall2015,Yan2022}. 

We choose the DVR basis in a box on a Cartesian grid of points. The method has two convergence parameters (for each Cartesian direction) associated with the basis choice: the size of the box, and the grid spacing. 
We use a DVR grid point spacing $(0.15, 0.15, 0.36)w_x$ and set the distance from the outermost trap centers to the system boundaries to $L_0=(3,3,7.2)w_x$. 
An assessment of the convergence is given in App.~\ref{sec:convergence}. We find that the Hubbard parameters presented in this paper are likely converged to a relative accuracy of $10^{-8}$ or better.

In our calculations, we make use of spatial reflection symmetries in $x$, $y$, and $z$ directions.
The lattice is on the $z=0$ plane so it's $z$-reflection symmetric, and the coordinate origin is set such that the lattices presented in the paper have $x$ and $y$ direction reflection symmetries. Unless otherwise mentioned, we calculate the $z$-reflection-even sector only to find the lowest energy states. For 1D lattices, the lattices sites are on $y=z=0$ line, so we further only consider the $y$- and $z$-reflection-even sectors.

\subsection{Maximally localized Wannier functions}
\label{sec:MLWF}
\def\wfi{W_{i\mu}}
\def\wfj{W_{j\mu}}
\def\dmia{D^{(\mu)}_{ia}}

Given the eigenstates of the DVR calculation obtained by the methods of the last section, we now describe how to obtain the MLWF, essentially following the techniques in Ref.~\cite{MLWF}. We truncate the eigenstates to those in the ``bands" of interest. For a single-band model, it suffices to truncate to a number of eigenstates equal to the number of traps. In the present paper, we truncate by keeping the lowest-energy states, but other truncation schemes could be used. 

In this truncated eigenbasis $\{\ket{\phi_a}\}_\mu$ chosen for the $\mu$-th band, we seek a new maximally-localized basis
\begin{equation}
\ket{W_{i\mu}}=\sum_{a\in A_\mu}D^{(\mu)}_{ia}\ket{\phi_a},
\end{equation}
where $i$ indexes sites, $D^{(\mu)}_{ia}$ is a unitary matrix within $\mu$-th band, and $A_\mu$ is the low-energy subset of the eigenbasis picked to form the $\mu$-th band. The $D^{(\mu)}_{ia}$ is chosen to minimize the sum of spatial spreads of the Wannier functions,  
\begin{equation}
    \Omega_\mu=\sum_i \left[ \bra{\wf} r^2 \ket{\wf} - \sum_{\xi} \bra{\wf} r_\xi \ket{\wf}^2 \right],
\end{equation}
where $r_\xi=x,y,z$.
Other similar cost functions can be used, as described in Ref.~\cite{MLWF}, but all should give equivalent results when the Hubbard-regime approximations described above are valid. 
For numerical purposes, it is convenient to work with an equivalent form obtained by subtracting the unitarily invariant part $\sum_{\xi}\mathrm{tr}_i \left[P_\mu r_\xi(\mathrm{1}-P_\mu)r_\xi\right]$, where the trace $\mathrm{tr}_i$ is over all $\wfi$ for the given $\mu$, and $P_\mu=\sum_i\ket{\wfi}\bra{\wfi}$ is the projector onto the $\mu$-th band,
\begin{align}
    \tilde{\Omega}_\mu
    =\blank \sum_{i\neq j}\sum_{\xi} |\bra{\wfi} r_\xi \ket{\wfj}|^2 \nonumber\\
    =\blank \mathrm{tr}_i\left[X'^2+Y'^2+Z'^2\right], \label{eq:localization-cost}
\end{align}
where pure off-diagonal matrices $X'$, $Y'$ and $Z'$ are defined as, for example, $X'_{ii}=0$ and $X'_{i\neq j}=\bra{\wfi} x \ket{\wfj}$. It is worth noting that the cost function is always non-negative, and zero value is taken if all the matrices $X'$, $Y'$ and $Z'$ are diagonalized simultaneously. For 1D since only $X'$ is nonzero, we make use of this fact and perform eigendecomposition to derive $D^{(\mu)}_{ia}$ for better performance. In higher dimensions, these matrices don't in general commute in the DVR basis, since it is an incomplete basis, and therefore numerical minimization is needed. But as shown in App.~\ref{sec:proveReal}, it can be proven that the $\tilde{\Omega}_\mu$ doesn't depend on phase of matrix elements of $D^{(\mu)}_{ia}$, so the latter can be reduced to orthogonal matrix. The minimization over orthogonal $D^{(\mu)}_{ia}$ for each $\mu$ can be done by established Riemannian manifold optimization algorithms, implemented by the \emph{pymanopt} package \cite{Pymanopt}.

After constructing MLWF by obtaining $D^{(\mu)}_{ia}$ for every band, we can now directly calculate Hubbard parameters $t$ and $V$ from \cref{eq:t,eq:V}. These equations give
\begin{align}
    V_{i\mu} =\blank \sum_a \epsilon_a |D^{(\mu)}_{ia}|^2, \label{eq:Vwf}\\
    t_{ij\mu} =\blank -\sum_a \epsilon_a (D^{(\mu)}_{ia})^*D^{(\mu)}_{ja} & i\neq j, \label{eq:twf}
\end{align}
and 
\begin{align}
    U_{i;\mu\nu\delta\sigma} = \blank
    \frac{4\pi\hbar^2a_s}{m} \sum_{abcd} (D^{(\mu)}_{ia} D^{(\nu)}_{ib})^* D^{(\delta)}_{ic} D^{(\sigma)}_{id} \nonumber\\
    \blank\times \int \! d^d\bmr\, \phi^*_a(\bmr) \phi^*_b(\bmr) \phi_c(\bmr) \phi_d(\bmr),
\end{align}
where $\epsilon_a$ is the energy of the single-particle eigenstate $\ket{\phi_a}$, and the integral is done numerically.

Since $\sum_{i\mu} V_{i\mu}$ is irrelevant in particle number conserving system, in practice we shift $V_{i\mu}$ to the average value over all $V_{i1}$'s in the lowest $\mu=1$ band.

\subsection{Results}
\label{sec:HubbardParamResults}

This section presents the results of calculating Hubbard parameters in various 1D and 2D lattices. 
For concreteness, the trap parameters are those of recent experiments with $^6$Li atoms~\cite{Spar2022,Yan2022}, but all of our methods are general. For rectangular lattices, the lattice constant is set to $a_x=1550\unit{nm}$ along the $x$ direction and $a_y=1600\unit{nm}$ in the $y$ direction
\footnote{Utilizing anisotropic lattice constants allows experiments to avoid low-frequency beatings at diagonal sites \cite{Yan2022}.}.
For triangular-type lattices e.g. kagome and honeycomb with defects, we use $a=1550\unit{nm}$ for all neighbor bonds. We use a $\lambda=780\unit{nm}$ laser wavelength, and use an isotropic trap waist $w=w_x=w_y=1000\unit{nm}$ to generate circular traps with depth $V_0=h\times 52.26\unit{
kHz}$ ($h$ is Planck's constant). This results in an isotropic Rayleigh range in $z$ direction of $z_R=\frac{\pi w^2}{\lambda}=4028\unit{nm}$. The $s$-wave scattering length is $a_s=93.66\unit{nm}$. These are the parameters used in our calculations throughout the paper unless mentioned otherwise.

\begin{figure*}[ht]
    \centering
    \includegraphics[width=\textwidth]{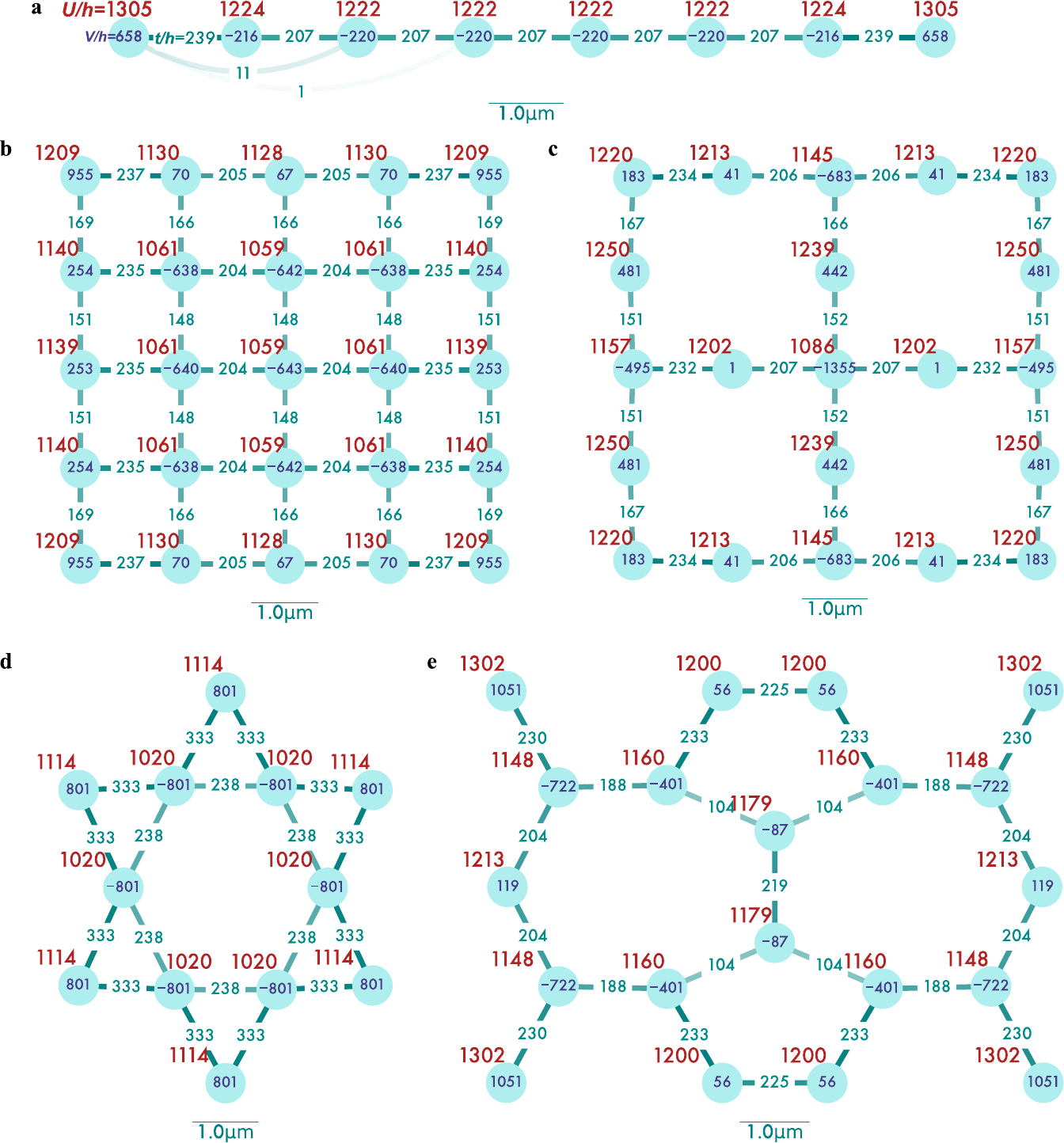}
    \caption{Hubbard parameters calculated for  
    (a) an $8$-site chain,
    (b) a $5\times5$ rectangle,
    (c) $2\times2$ Lieb plaquettes,
    (d) a star of David (kagome plaquette), and
    (e) a  honeycomb lattice with a Stone-Wales defect.
    Each node of the graph shown is centered at the Wannier function peak position. Numbers offset from the nodes are on-site interactions $U$, inside circles are on-site potentials $V$, and on bonds are tunnelings $t$, all in units of $h\times\unit{Hz}$. For each geometry, we shift the $V_i$ so that their average is zero, since only relative differences between $V_i$'s have physical meaning. The transparency of bonds reflects the tunneling strength.
    Further neighbor tunnelings are showed in (a), but omitted elsewhere. Tweezer parameters are specified in the main text.}
    \label{fig:HubbardParam}
\end{figure*}

\fig{HubbardParam} shows the Hubbard parameters calculated for five example geometries in the single-band limit.
Figs.~\ref{fig:HubbardParam}(a) and (b)  show chains and rectangular lattices, small versions of geometries routinely realized in optical lattices. Figs.~\ref{fig:HubbardParam}(c) and (d) show geometries that are  more challenging, but possible, to realize in an optical lattice: a $2\times2$ Lieb lattice~\cite{Taie2015} and a kagome lattice~\cite{kagomePRL2012,melchner2022quantum}. Finally, \fig{HubbardParam}(e) shows a geometry that is straightforward to create in tweezers that would be extraordinarily difficult or impossible to realize in an optical lattice, a honeycomb lattice with a Stone-Wales defect \cite{defectHoneycomb}. For all the calculations the manifold optimizations (and matrix diagonalization for \fig{HubbardParam}(a)) converge to machine precision for the resulting MLWFs. These calculations each run in no more than around 1 minutes on a laptop. The majority of the time is spent on diagonalizing the DVR matrices.

In addition to demonstrating the efficacy of the DVR$+$MLWF method described in \sect{effModel}--\ref{sec:MLWF}, these results allow us to assess the accuracy of range-truncating the tunneling and interactions in Eq.~\eqref{eq:FH}.
\fig{HubbardParam}(a) shows that the tunneling amplitudes decay rapidly with increasing distance for the 1D chain.
The same is true for the $U_{ijkl}$'s, where we find the on-site $U_{i}$'s are two orders of magnitude larger than any $U$'s involving two or more sites. These justify the effective Hubbard model description.


\begin{figure*}[htb]
    \centering
    \includegraphics[width=0.8\textwidth]{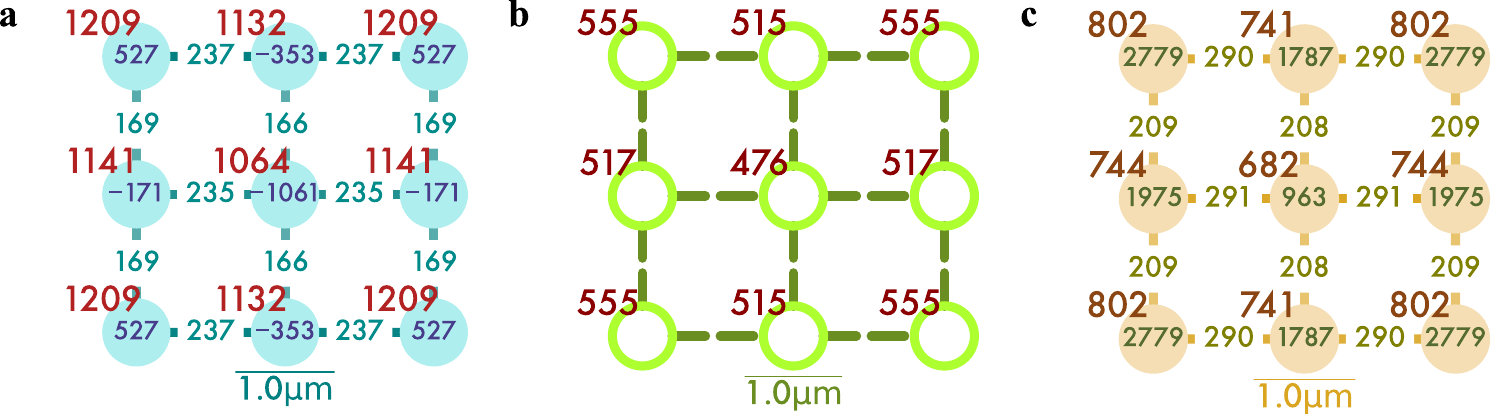}
    \caption{Two-band Hubbard parameters for (a) the first band, (b)  interband interactions between the first band and the $z$-excited band, and (c) the $z$-excited band.  The $z$-direction is out of the plane, and the $z$-excited band is the lowest energy excited band for the parameters considered, as described in the main text.}
    \label{fig:multiband}
\end{figure*}

Higher bands can be included in the Hubbard models if desired, and the methods employed here work without modification. \fig{multiband} illustrates this for two-band Hubbard parameters for a $3\times3$ rectangular lattice consisting of two lowest-energy bands: the first spanned by MLWFs that are approximately single-trap ground states, and the second spanned by the first excited state, which is odd under $z$-reflections. Given the reflection symmetries of two bands, the only non-zero interband $U_{\mu\nu\delta\sigma}$'s are $U_{\mu\nu\nu\mu}$, $U_{\mu\mu\nu\nu}$ with $\mu\neq\nu$, and they are equal because all MLWFs are real. This result shows the power of the DVR+MLWF calculation method. Access to higher bands could pave the way to simulating multi-orbital models, an important aspect of real materials.

An interesting feature revealed by  \fig{HubbardParam} is that  Hubbard parameters vary spatially over the  tweezer arrays, despite these tweezers'  equal spacing and depth, due to the cross-talk illustrated in Fig.~\ref{fig:Tweezer}(c). This shows the need to adjust trap parameters to achieve uniform $t$, $U$, and $V$, a task we take up in \sect{equalizeHubbard}.
We note, however, that the non-uniformity is largely concentrated near the edge of the system, as is shown in \fig{HubbardParam}(a) and (b), inspiring one of our protocols to equalize Hubbard parameters discussed in detail in \sect{equalizeHubbard}.

\section{Equalizing Hubbard parameters}
\label{sec:equalizeHubbard}

For quantum simulations often we are interested in translation-invariant models, but \sect{HubbardParam} showed that equally-spaced and equal-depth traps do not lead to spatially uniform Hubbard parameters. This is natural, because optical tweezer arrays, in contrast to optical lattices, lack translation invariance. The cross-talk between traps breaks the trap uniformity, as is shown in \fig{Tweezer}(c). This effect is particularly severe at the boundaries.
So it is an important question of whether and how trap parameters can be chosen to equalize the Hubbard parameters.



\subsection{Method}

One approach to equalizing the trap parameters is to introduce a cost function measuring how unequal the parameters are and to use optimization algorithms to minimize this over the space of trap parameters: the trap center positions $\bmr_i$
and the individual trap depths $V_{0i}$.
We use the cost function
\begin{equation}\label{eq:cost}
    C=\frac{1}{\tilde{t}^2}\sum_q\frac{1}{N_q}\sum_i \left(q_i-\tilde{q}\right)^2
\end{equation}
where $q=t_x, t_y, V$, and $U$ labels the Hubbard parameters being equalized, $\{\tilde{q}\}$ are target values for the parameters (we allow separate tunnelings $t_x$ and $t_y$ for anisotropic lattices), $i$ indexes sites for $U$ and $V$ and indexes bonds for $t_x$ and $t_y$, $N_q$ labels the number of Hubbard parameters of kind $q$, and $\tilde{t}$ is the smaller of $\tilde{t}_x$ and $\tilde{t}_y$. This characterizes the parameter fluctuations compared to the most sensitive energy scale in the model.
The optimization algorithms we use are the sequential least-square programming (SLSQP) method in \emph{scipy} package and principal axis (PRAXIS) method in \emph{nlopt} package in Python programming language. We apply both methods and choose the better minimum.

This cost function depends on the target Hubbard parameters $\{\tilde{q}\}$. One cannot simply choose these arbitrarily, as there are constraints, for example on how large parameters such as the tunnelings can be. For example, for two traps, there is a maximum ``tunneling" when the two traps are focused at the same location. In practice, in order to stay in the Hubbard regime, the tunneling needs to be set much below this maximum. Additionally, choosing target values for which a known trap configuration gives Hubbard parameters reasonably close to the target values will make the optimization more efficient, reducing the number of Hubbard parameter calculations required and reducing the chances of getting stuck in a local, rather than global, minimum.

To obtain target $\{\tilde{q}\}$, we have found the following procedure works well. We first calculate Hubbard parameters for initial trap configurations in the geometry desired, and then we choose the largest interaction in the lattice as our target $\tilde{U}$, and the smallest tunneling in the $x$ and $y$ directions as our target $\tilde{t}_x$, $\tilde{t}_y$. To achieve the desired $U/t_a$ values we can scale the geometry.
The target value for $V$ is set to zero (recall that mentioned in \sect{MLWF} the zero of $V$ is shifted in every calculation to the average of the lowest band).

In the following, we present results of equalizing Hubbard parameters to uniform target values. We will present results from three protocols: the first two use only capabilities demonstrated in existing experiments: adjust only the trap positions $\bmr_{i}$ and depths $V_{0i}$, while the third also allows the tweezer waists $w_i$ to be tuned, and we suggest a method to accomplish this.

\subsection{Equalization with previously demonstrated tweezer techniques}

\begin{figure*}[ht]
    \centering
    \includegraphics[width=0.8\textwidth]{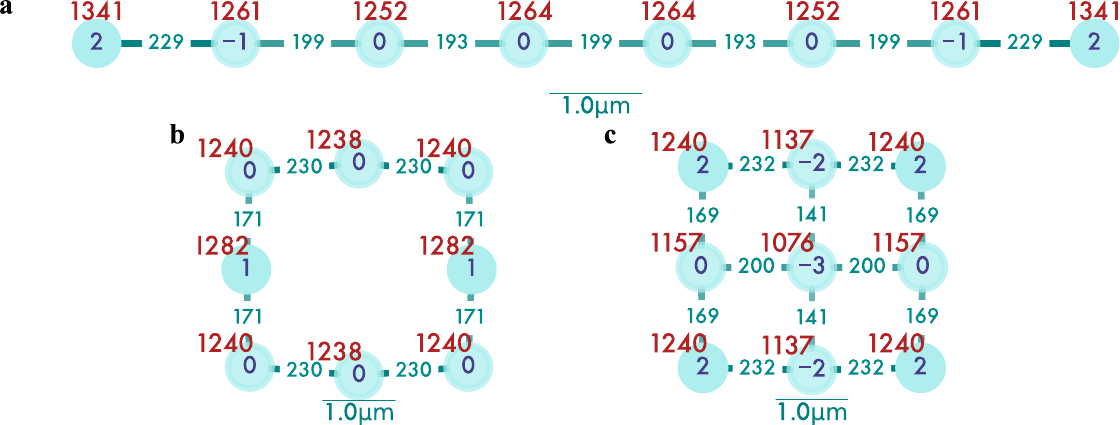}
    \caption{Hubbard parameters for trap parameters that give the closest-to-equalized Hubbard parameters [according to Eq.~\eqref{eq:cost}] achievable with by varying trap position and depth for
    (a) an $8$-site chain,
    (b) one Lieb plaquette, and
    (c) a $3\times3$ rectangular lattice.
    Number labels are as in Fig.~\ref{fig:HubbardParam}.}
    \label{fig:equalizeCurrent}
\end{figure*}

We first try to equalize Hubbard parameters based on how previous experiments \cite{Spar2022,Yan2022} were performed: we adjust the trap positions $\bmr_{i}$ and depths $V_{0i}$. The results are shown in \fig{equalizeCurrent}. In this figure, especially \fig{equalizeCurrent}(b), we can see we can see how equalizing Hubbard parameters requires distorting the experimental trap parameters. In \fig{equalizeCurrent} (and following figures), the positions of the nodes are located at the peaks of the Wannier functions; the transparency of nodes shows the relative depths of trap $V_{0i}$ compared to the trap with maximum depth.

From \fig{equalizeCurrent}, we can see that although the total variation of $V$ over the lattice is suppressed to less than than $\sim h\times 5\unit{Hz}$, the differences for interaction $U$ and tunneling $t$ are only modestly improved compared to the uniform trap configurations, such as results in Figs.~\ref{fig:HubbardParam} and \ref{fig:multiband}. (The squared variation is reduced by a few percent, while the maximum difference can even increase slightly.) This is particularly severe if we think of the most sensitive energy scale $\tilde{t}$ in Hubbard model, as the maximal fluctuation of $U$ is as large as $\tilde{t}$. The variation is the most significant when comparing between boundary and bulk sites. And comparing among various lattices, it is more severe in lattices with more neighbors.

It is natural that not all parameters can be equalized since the number of Hubbard parameters is larger than the number of tunable trap parameters. For example, a large anisotropic square lattice has roughly four Hubbard parameters per trap ($t_x$, $t_y$, $U$, and $V$), but only three parameters to tune per trap ($x$ and $y$ location, and trap depth $V_{0i}$). Tradeoffs can be made in the accuracies of the $t_{x,y}$, $V$, and $U$ by reweighting the terms in the cost function \eq{cost}, but at best one can equalize two parameters fairly accurately.

In order to fully equalize all the Hubbard parameters, more degrees of freedom must be introduced. Here we present two proposals to achieve that goal: the first uses extra ``ghost traps" on the edges, and uses only tuning of trap parameters already demonstrated in experiment (positions and depths), and the second uses ``trap-dependent waists''. The ``ghost traps'' proposal is the more feasible approach with current experimental capabilities, although still must be handled with care (keeping ghost traps sufficiently distinct to avoid beating, and requiring slightly more laser power to generate ghost trap layers). The ``trap-dependent waist'' proposal is more speculative, and has challenges to obtain sufficient flexibility in tuning the waists with AOMs or stability of SLMs, but which nevertheless conceivably could be developed into a useful tool.

\begin{figure*}
    \centering
    \includegraphics[width=0.9\textwidth]{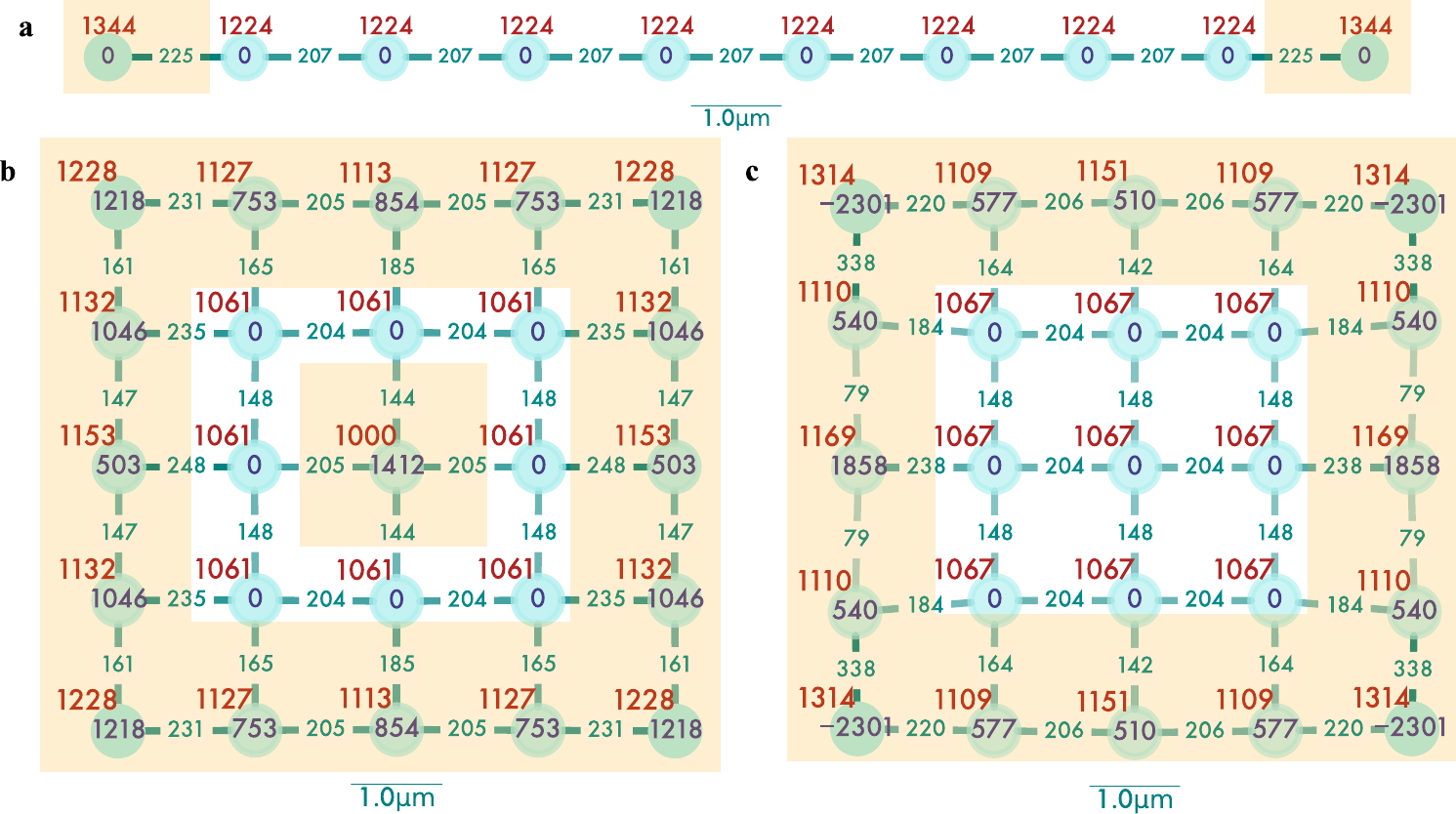}
    \caption{Hubbard parameters equalized as in Fig.~\ref{fig:equalizeCurrent}, augmented with ``ghost sites" on the boundaries for
    (a) $8$-site chain, (b) one Lieb plaquette and (c) a $3\times3$ rectangular lattice.
    Number labels are as in Fig.~\ref{fig:HubbardParam}.
    \label{fig:ghostSite}}
\end{figure*}

\subsection{Proposal 1 for improved equalization: ghost sites}

The first proposal to equalize Hubbard parameters is to add ``ghost sites'', as shown in \fig{Tweezer}(d). Inspired by the observation that the Hubbard parameter non-uniformity is mostly near system edges, an experiment can simply restrict to measurements on sites away from the edges, and obtain good uniformity. If, additionally, one optimizes for uniformity only on sites not along the edge by including only non-edge sites in the cost-function, the parameters can be made uniform (we obtain a $2\times 10^{-4}$ relative variation between parameters in the example, with error that is set by the optimization algorithm tolerance, rather than a fundamental limitation of the protocol). This is illustrated for the chain and square lattices in Figs.~\ref{fig:ghostSite}(a) and (c). This protocol is suitable when one is studying lattices of simple patterns, such as 1D chain or 2D square lattices. Then physical quantities such as the particle number in the edge sites (ghost traps) are not included in measurements of observables for the purposes of quantum simulation. This way, the edges can be seen as a special boundary condition to the rest of the tweezer array.

A slightly more difficult case in which to employ ghost traps is for a lattice such as the Lieb plaquette, as several boundaries occur not as the edges of a large array, but within each unit cell. Another example is when one wants to study a finite-size Hubbard system with sharp edges. In this case, ghost traps can still be used, but one must ensure that particles do not occupy the ghost traps. This can be done by ensuring the on-site potential on the ghost traps is large enough to prevent occupation, which is done by giving the equalization algorithm a $5\%\times V_0\approx h\times 2500\unit{Hz}$ smaller initial value for $V_0$ on the center ghost trap, which is about one order of magnitude larger than tunnelings. Fig.~\ref{fig:ghostSite}(b) shows that this approach again works well for the Lieb plaquette by applying the special initial guess the central ghost trap ``inside" the plaquette. This relative error is better than $2\times10^{-4}$ and controlled by optimization convergence criterion, with a central ghost trap having $V_i=h\times 1400\unit{Hz}$.


\subsection{Proposal 2 for improved equalization: tunable tweezer trap waists}\label{sec:waist}

Adding ghost traps is effective, but requires sacrificing many traps (and associated laser power and imaging region) to traps that are not being measured. Ghost traps are lattice geometry specific, and it might not work well for irregular lattice geometries, such as a lattice with a Stone-Wales defect as shown in \fig{HubbardParam}(e). Additionally, although we don't explore it here, it would not provide any significant ability to engineer higher bands' Hubbard parameters. In this section, we introduce a second method that offers more tuning parameters and does not require additional traps.

This second proposal is to adjust each tweezer trap waist $w_i$ individually, as is shown in \fig{Tweezer}(a), which increases the number of trap parameters and gives a natural way to tune the $U_i$, $V_i$, and $t_{ij}$ more independently. Although site-dependent waists have not been demonstrated in experiment, one idea to change the width of each trap individually is to program it using the same stroboscopic averaging that is used to create the tweezers in the first place \cite{Yan2022}. There may be limited control of the exact shape created due to imaging resolution and rate of strobing, but some control on trap width should be possible. 

\begin{figure*}
    \centering
    \includegraphics[width=0.8\textwidth]{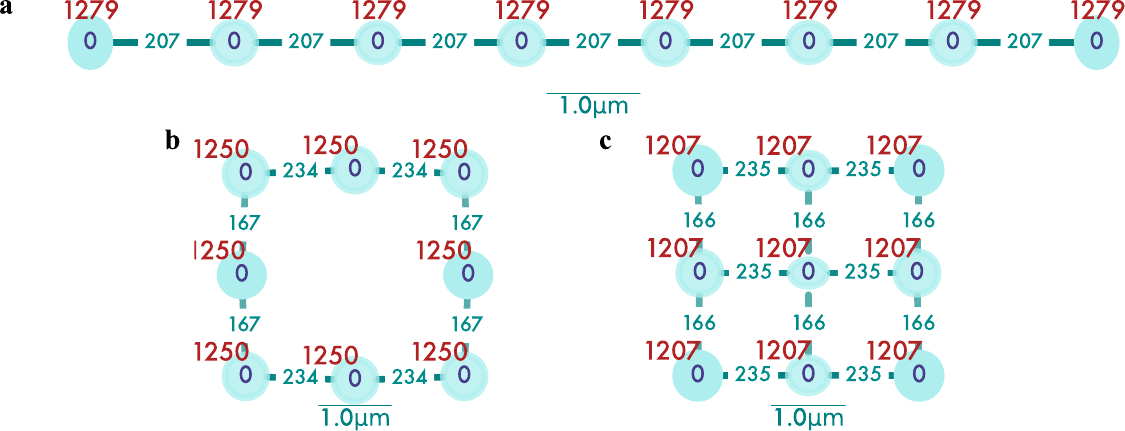}
    \caption{Hubbard parameters equalized as in Fig.~\ref{fig:equalizeCurrent} and also allowing waists to vary:
    (a) an $8$-site chain allowing  $x$-direction waist to vary, and (b) a Lieb plaquette and
    (c) a $3\times3$ rectangular lattice allowing trap waists to vary in the $x$- and $y$-directions.
    Shape of each node shows the ratio between distorted trap waists in $x$ and $y$ directions.
    Number labels are as in Fig.~\ref{fig:HubbardParam}.}
    \label{fig:equalizeWaist}
\end{figure*}

The Hubbard parameters after optimizing using these additional degrees of freedom are shown in \fig{equalizeWaist}. We vary only waists $w_{i,x}$ in the $x$ direction in \fig{equalizeWaist}(a)'s 1D chain, while for the more complicated 2D systems shown in \fig{equalizeWaist}(b) and (c), we allow for both $x$ and $y$ direction waists $w_{i,x}$, $w_{i,y}$ to vary.


\section{Conclusion}
\label{sec:conclusion}

We have calculated the effective Hubbard model parameters for programmable fermionic optical tweezer arrays, a novel platform for quantum simulations and computation. We used a combination of a DVR method to obtain single-particle eigenstates, truncating these to the target energy band, and optimizing over the manifold of unitary transformations  for that energy subspace to obtain the MLWF. We found that this method allowed us to efficiently calculate Hubbard parameters. We also presented results for several geometries: finite 1D chains, 2D square lattices, Lieb lattices, and hexagonal lattices with Stone-Wales defects. Multi-band calculations were demonstrated, for a two-band square lattice.
We note that the method is immediately applicable to other systems, including bosonic atoms and long-ranged interacting systems, such as lanthanide atoms~\cite{NorciaLanthanide2021,LiLanthanide2017,LiLanthanide2018}, polar molecules~\cite{KaufmanNi2021,Gorshkov2011,Yan2013,HazzardPolar2013,NataleBoson2019}, or Rydberg-dressed atoms~\cite{DefenuLRI2021,JohnsonRydbergDress2010,ZhouRydbergDressBoson2020}. In addition, our methods are equally effective for recently developing hybrid lattice-tweezer architecturesm, such as the optical lattices with programmable site-block tweezer beams \cite{Wei2023}, and optical quasicrystal realized by a set of incommensurate wavelength laser beams \cite{OpticalQuasicrystal2019,GottlobQuasicrystal2023}.

One feature revealed by our results is that even though a tweezer geometry may be uniform (equal spacings and depths for all traps), the resulting Hubbard parameters may be non-uniform. This effect is most pronounced near the edges of the trap arrays. 

Since one often wants to study spatially uniform Hubbard models, we introduced and demonstrated  protocols to design trap array parameters to achieve this.
We expect that optimization with an appropriately chosen cost function should also allow one to determine trap parameters that lead to desired non-uniform Hubbard models.
The enhancement of programmability of tweezer arrays can further contribute to the realization of the proposed fermionic gate design and control \cite{FermionGate2023,FermionVQE2023}.




Although published experiments with tunnel-coupled tweezers~\cite{Spar2022,Yan2022,FermionDoubleWell,BosonDoubleWell} have worked with modest-sized systems having at most $\sim 50$ traps, experiments with hundreds or perhaps thousands of sites and atoms seem possible by extending current techniques. Experiments with tweezer arrays of Rydberg atoms now routinely work with over a hundred atoms. For example, recent experiments have reached 225 sites to achieve a perfect filling of one atom per trap~\cite{Tian2023} over 33\% of the time in a room-temperature apparatus. Cryogenic experiments can offer even larger arrays \cite{Schymik_crygenic2022}. And there are clear routes to increasing this number. There are no fundamental obstacles to applying the same techniques to tunnel-coupled tweezers. 
Our code readily handles such cases, scaling well with the number of traps $N_{\text{trap}}$.
The time and memory required for DVR calculations scale no worse than $N_{\text{trap}}^2$. The time and memory costs to find the unitary basis transformation are negligible and scale no worse than DVR.
In practice, we are able to calculate Hubbard parameters of systems with 100 traps in about an hour on a laptop.

\acknowledgements
We acknowledge Waseem Bakr, Zoe Yan, Benjamin Spar, and Max Prichard for many useful conversations, especially for suggesting the basic idea of ghost traps. We acknowledge support from the Robert A. Welch Foundation (C-1872), the National Science Foundation (PHY-1848304 and CMMI-2037545), the Office of Naval Research (N00014-20-1-2695), and the W. F. Keck Foundation (Grant No. 995764). Computing resources were supported in part by the Big-Data Private-Cloud Research Cyberinfrastructure MRI-award funded by NSF under grant CNS-1338099 and by Rice University's Center for Research Computing (CRC). E.I.G.P. acknowledges support by the grant DE-SC-0022311, funded by the U.S. Department of Energy, Office of Science. K.H.'s contribution benefited from discussions at the Aspen Center for Physics, supported by the National Science Foundation grant PHY-1066293, and the KITP, which was supported in part by the National Science Foundation under Grant No. NSF PHY-1748958.

\appendix


\section{DVR convergence }
\label{sec:convergence}

\begin{figure*}
    \centering
    \includegraphics[width=0.95\textwidth]{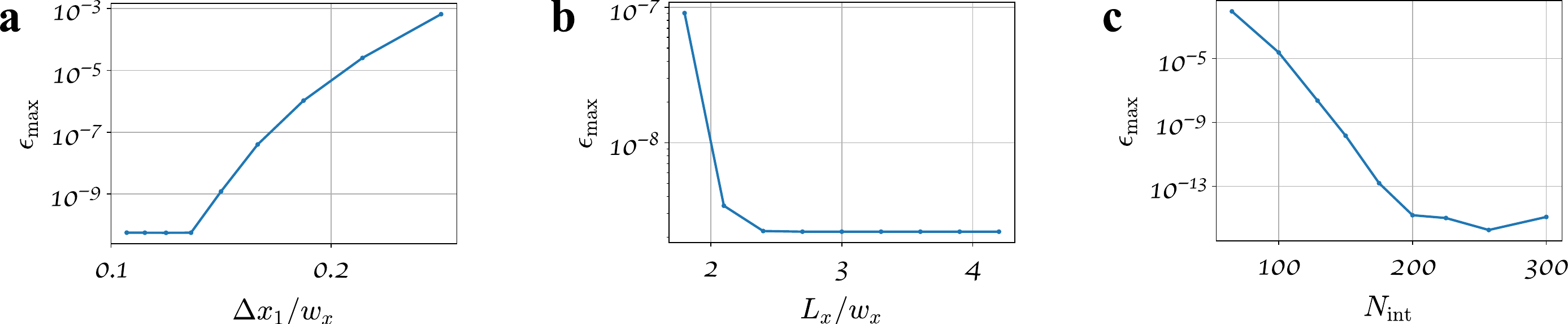}
    \caption{The maximal error of $8$-site chain Hubbard parameters vs DVR grid spacing $\Delta x=(1, 1, 2.4)\times\Delta x_1$, grid range $L_0=(1,1,2.4)\times L_x$ and numerical integration grid point number $N_\mathrm{int}$. Definition of each quantity is explained in App.~\ref{sec:convergence}.}
    \label{fig:maxErr}
\end{figure*}

In the main text, we use the DVR grid point spacing $\Delta x=(0.15, 0.15, 0.36)w_x$ and set the box size by setting  the distance from outermost trap centers to the system boundaries to $L_0=(3,3,7.2)w_x$. This was shown to be highly converged for a single trap in Ref.~\cite{Yan2022}. We additionally check the convergence in an $8$-site chain Hubbard parameter calculation, by comparing values of $t$, $U$, and $V$ among different grid point spacings $\Delta x$ and ranges $L_0$. In addition, we also check the $U$ numerical integration convergence by varying the number of points in the integration grid. All above results are summarized in \fig{maxErr}, indicating good convergence for the results presented in the main text. Based on these results, we expect worst-case errors from all sources across all results in the text to be $10^{-8}$ or smaller. In the remainder of this appendix, we will describe how the convergence of the algorithm was assessed.

We characterize the error of a given Hubbard parameter by extrapolating the convergence parameter (e.g. grid spacing $\Delta x$ or spatial buffer size $L_0$) to its fully converged value (e.g. $\Delta x\to 0$ or $L_0\to\infty$), and then calculating its maximum relative error over the whole geometry $\epsilon_\mathrm{max}$, defining the error as the relative difference between the grid being assessed and the extrapolated grid:
\begin{equation}
    \epsilon_\mathrm{max}=\max_i\frac{| q_i^{(\text{choice})} - q_i^{*} |}{ | q_i^{*}|}.
\end{equation}
Here $q_i$ can be any Hubbard parameter $t_{ij}$, $V_i$ or $U_i$ and the site (bond) index $i$ is over the entire lattice. $q_i^{(\text{choice})}$ means $q_i$ is calculated for some convergence parameters, and $q_i^*$ is the $q_i$ extrapolated at the infinitely fine grid.

The extrapolation is performed by fitting the 3 most accurate grid choices for the quantity $q_i$ using the exponential function:
\begin{equation}
    q_i^{(\text{choice})}(N)=ae^{-bN}+c,
\end{equation}
where $N$ is the grid fineness parameter. It can be $N_{\Delta x}$, $N_L$ or $N_\mathrm{int}$ whose definitions are given below. The extrapolation is performed to $N\to\infty$, so that $q_i^{*}=c$ for each fit.

\fig{maxErr}(a) shows the convergence with the DVR grid spacing $\Delta x$. We consider results for grid spacings $\Delta x=L_0/N_{\Delta x}$ with $N_{\Delta x}=12,14,\ldots,28$, and we fix $L_0=(3,3,7.2)w_x$. These spacings range from $\Delta x^*=(0.107, 0.107, 0.0.257)w_x$ to $(0.25, 0.25, 0.6)w_x$. The grid spacing we use in the main text corresponds to $N_{\Delta x}=20$. These results suggest the grid spacing convergence error is as small as $10^{-9}$ for calculations in the main text.

\fig{maxErr}(b) shows the convergence with the total size of the DVR box, measured by $L_0$. We vary $L_0$ over $L_0 = (0.15, 0.15, 0.36)w_x \times N_L$ where $N_L=12,14,\ldots,28$ and we fix $\Delta x=(0.15,0.15,0.36)w_x$, with the best convergence at $L_0^*=(4.2, 4.2, 10.08)w_x$. The box size we use in the main text is $N_L=20$. These results show that the box-size convergence error is on the order of $10^{-9}$. We see this error saturates to the error from $\Delta x$.

Finally, \fig{maxErr}(c) shows the convergence of $U$ with the numerical integration grid density. We choose the numerical integration region to extend to $1.2$ times the DVR box, and the integration grid spacing to be set by the number of grid points in each dimension, $N_\mathrm{int}$. In the main text, we use $N_\mathrm{iint}=257$. The results show that the numerical integration converges to the order of machine precision.

For higher band case, we expect error convergence to be similarly exponential, but with a slight larger prefactor, according to calculations in \cite{Wall2015}.

In summary, all of our Hubbard parameter systematic numerical errors are under control to an error less than $10^{-8}$. And our parameter choices are close to saturation.

\section{Proof on all-real MLWF transformation matrix entries}
\label{sec:proveReal}
\def\dia{D_{ia}}

In this section, we prove that in our case, all entries in MLWF unitary transformation matrix $\dmia$ can be chosen to be real. This means it can be reduced to the special orthogonal matrix. This is from the fact that the tweezer trapping continuum Hamiltonian is real. Without loss of generality, we suppress the band index $\mu$ for the derivation.

Our starting point is the cost function \eq{localization-cost} to minimize in the main text in order to find out MLWFs:
\begin{align}
    \tilde{\Omega}
    =\blank \sum_{\xi}\sum_{i\neq j} |\bra{W_i} r_\xi \ket{W_j}|^2 \nonumber\\
    =\blank \sum_\xi\sum_{i\neq j} |\sum_{ab} (D_{ia})^*D_{jb} R_{\xi,ab}|^2,
    \label{eq:Omega}
\end{align}
where $R_{\xi,ab}\equiv\bra{\phi_a} r_\xi \ket{\phi_b}$ is shorthand for matrix elements of position operators in the eigenbasis $\{\ket{\phi_a}\}$. 

Since the continuum Hamiltonian is real, it can be proven that the eigenstates can be all-real. Think of the complex conjugate of the eigenequation of $\phi_a$ with eigenenergy $\epsilon_a$:
\begin{equation}
    H^*\phi^*_a = \epsilon^*_a \phi^*_a \to H\phi^*_a = \epsilon_a \phi^*_a.
\end{equation}
Because both $H$ and $\epsilon_a$ are real, $\phi^*_a$ is the degenerate eigenstate of the same eigenenergy. Proper linear combinations of $\phi_a$ and $\phi^*_a$ derive two real solutions. Therefore, we claim that $R_{\xi,ab}$ matrices are hermitian and real.

As mentioned in the main text, this function is non-negative, and the zero value can be taken if $R_\xi=X,Y,Z$ can be simultaneously diagonalized. In a 1D lattice, only $R_\xi=X$ matrix would be non-zero due to reflection symmetry, and the minimum of $\tilde{\Omega}$ is found at the $\dia$ diagonalizing $X_{ab}$. This $\dia$ is orthogonal (has only real entries) because eigenstates of the real hermitian matrices $R_{\xi,ab}$ can be chosen to be real, with the reason illustrated above.

For higher dimensions the $D_{ia}$ that minimize $\tilde{\Omega}$ are no longer the eigensolutions of $X_{ab}$, as the matrices $R_\xi$ are basis dependent and in general don't commute. The optimization needs to be done numerically. 

Nevertheless, the matrix $\dia$ can be simplified to be real entries by the following proof.

Since we cannot find a single real (orthogonal) matrix to diagonalize 3 $R_\xi$'s, we find 3 orthogonal matrices $O^\xi$ to diagonalize the real hermitian matrices $R_{\xi,ab}$ respectively:
\begin{equation}
    R_{\xi,ab} =\sum_{mn} r_{\xi,m} \delta_{mn} O^\xi_{am} O^\xi_{bn},
\end{equation}
where $r_{\xi,m}$ is the $m$-th eigenvalue of matrix $R_\xi$.

Then we decompose unitary matrix into two matrices $\dia=\sum_b O^\xi_{ab}Z^\xi_{bi}$, where we absorb all complex phases into $\xi$-dependent matrix $Z^\xi$. It is clear that $Z^\xi$ is unitary as well.

\begin{widetext}
For each $\xi$ we can expand the terms in cost function \eq{Omega}, and simplify the equation by substituting $O^\xi$:
\begin{align}
    \sum_{i\neq j} |\sum_{ab} (\dia)^*D_{jb} R_{\xi,ab}|^2
    =\blank \sum_{i\neq j} |\sum_{ab} (\dia)^*D_{jb} \sum_{mn} r_{\xi,m} \delta_{mn} O^\xi_{am} O^\xi_{bn}.|^2 \nonumber\\
    =\blank \sum_{i\neq j} |\sum_{ab} (\dia)^*D_{jb} \sum_{m} r_{\xi,m} O^\xi_{am} O^\xi_{bm}.|^2 \nonumber\\
    =\blank \sum_{i\neq j} \sum_{m} |(Z^\xi_{mi})^*Z^\xi_{mj} r_{\xi,m}|^2.
\end{align}
The above is because of the orthogonality of $O^\xi$. We then expand the complex modulus:
\begin{align}
    \sum_{i\neq j} \sum_{m} |(Z^\xi_{mi})^*Z^\xi_{mj} r_{\xi,m}|^2
    =\blank \sum_{i\neq j} \sum_{m} (Z^\xi_{mi})^*Z^\xi_{mj} r_{\xi,m} \sum_{n} (Z^\xi_{nj})^* Z^xi_{ni} r_{\xi,n} \nonumber\\
    =\blank \sum_{mn} r_{\xi,m}r_{\xi,n} \sum_{i\neq j}(Z^\xi_{mi}Z^\xi_{nj})^* Z^\xi_{ni}Z^\xi_{mj}.
\end{align}
Now we focus on the only complex part:
\begin{align}
    \sum_{i\neq j}(Z^\xi_{mi}Z^\xi_{nj})^* Z^\xi_{ni}Z^\xi_{mj}
    =\blank \left(\sum_{ij} - \sum_{i=j}\right) (Z^\xi_{mi}Z^\xi_{nj})^* Z^\xi_{ni}Z^\xi_{mj} \nonumber\\
    =\blank \sum_{i} (Z^\xi_{mi})^* Z^\xi_{ni} \sum_{j} (Z^\xi_{nj})^*Z^\xi_{mj} - \sum_{i} (Z^\xi_{mi}Z^\xi_{ni})^* Z^\xi_{mi}Z^\xi_{ni} \nonumber\\
    =\blank \delta_{mn} - \sum_{i} |Z^\xi_{mi}|^2|Z^\xi_{ni}|^2.
\end{align}
The last Kronecker delta is from that $Z^\xi$ is unitary, and $\delta^2_{mn}=\delta_{mn}$.

We can see that every factor in the expression of $\tilde{\Omega}$ is independent on the phase of any unitary matrix entry $Z^\xi_{mi}$. So it suffices to assume all entries of $\dia$ to be real.
\end{widetext}





\nocite{*}
\bibliography{Hubbard.bib}

\end{document}